\documentclass[PRL,preprint,showpacs,preprintnumbers,superscriptaddress]{revtex4}

\usepackage{graphicx}
\usepackage{dcolumn}
\usepackage{bm}
\usepackage{amsmath}
\usepackage{epsfig}

\normalsize

\begin{document}

\title{Multi-Photon, Multi-Level Dynamics in a Superconducting Persistent-Current Qubit}

\author{Y. Yu}
\affiliation{Department of Electrical Engineering and Computer Science \\
Massachusetts Institute of Technology, Cambridge MA 02139}
\author{W.~D. Oliver}
\affiliation{MIT Lincoln Laboratory, 244 Wood Street, Lexington,
MA 02420}
\author{J.~C. Lee}
\affiliation{Department of Electrical Engineering and Computer Science \\
Massachusetts Institute of Technology, Cambridge MA 02139}
\author{K.~K. Berggren}
\affiliation{Department of Electrical Engineering and Computer Science \\
Massachusetts Institute of Technology, Cambridge MA 02139}
\author{L.~S. Levitov}
\affiliation{Department of Physics, Massachusetts Institute of
Technology, Cambridge MA 02139}
\author{T.~P. Orlando}
\thanks{to whom correspondance should be addressed}
\affiliation{Department of Electrical Engineering and Computer Science \\
Massachusetts Institute of Technology, Cambridge MA 02139}

\date{\today}

\begin{abstract}
Single-, two-, and three-photon transitions were driven amongst
five quantum states of a niobium persistent-current qubit. A
multi-level energy-band diagram was extracted using microwave
spectroscopy, and avoided crossings were directly measured between
the third and fourth excited states. The energy relaxation times
between states connected by single-photon and multi-photon
transitions were approximately $30 - 100 $ $\mu s$. Three-photon
coherent oscillations were observed between the qubit ground and
fourth excited states with a decoherence time of approximately
$50$ ns.


\end{abstract}
\pacs{03.67.Lx,03.65.Yz,85.25.Cp,85.25.Dq}
\maketitle
\date{\today}

Superconducting Josephson junctions are devices that exhibit
quantum phenomena amongst their macroscopic degrees of
freedom~\cite{Leggett87a,Leggett87b}. Demonstrations of
macroscopic quantum tunneling and energy level quantization
provide a basis for utilizing superconducting devices in quantum
computing applications~\cite{Clarke88a}. Quantum-state
superposition~\cite{Friedman00a,Wal00a}, multi-qubit
spectroscopy~\cite{Berkley03a,Xu05a}, coherent temporal
oscillations~\cite{Nakamura99a,Vion02a,Yu02a,Martinis02a,Chiorescu03a}
and elements of coherent
control~\cite{Pashkin03a,Yamamoto03a,Mcdermott05a} have been
demonstrated with Josephson-based quantum bits~\cite{Makhlin01a}.

Qubit spectroscopy and coherent oscillations can be externally
driven in several regimes: by single-photon or multi-photon
transitions, with weak or strong driving amplitude, and in
two-level or multi-level systems~\cite{macro_meso_note}. Several
examples exist of single-photon oscillations in weakly-driven
quasi-two-level
systems~\cite{Nakamura99a,Vion02a,Yu02a,Martinis02a,Chiorescu03a}.
Nakamura et al. demonstrated single- and multi-photon quantum
coherence in a strongly-driven, two-level, charge-qubit
system~\cite{Nakamura}. Likewise, Claudon et al. observed
single-photon coherent oscillations in a multi-level DC SQUID
potential well~\cite{Claudon04a}. Recently, Saito et al. reported
single- and multi-photon spectroscopy in a weakly-driven two-level
persistent-current (PC) qubit~\cite{Saito04a}. Most demonstrations
have utilized aluminum-based devices.

In this paper, we investigate \emph{multi-photon},
\emph{multi-level} spectroscopy and dynamics in a weakly-driven
niobium persistent-current qubit. We utilize static
and time-dependent spectroscopic information from single-, two-,
and three-photon transitions to plot a multi-level energy band
diagram of the qubit, to identify avoided crossings between its
macroscopic states, to measure the energy relaxation time $T_1$
from selected excited states, and to demonstrate three-photon
coherent oscillations between the qubit's ground state and fourth
excited state.

\begin{figure}
 \includegraphics{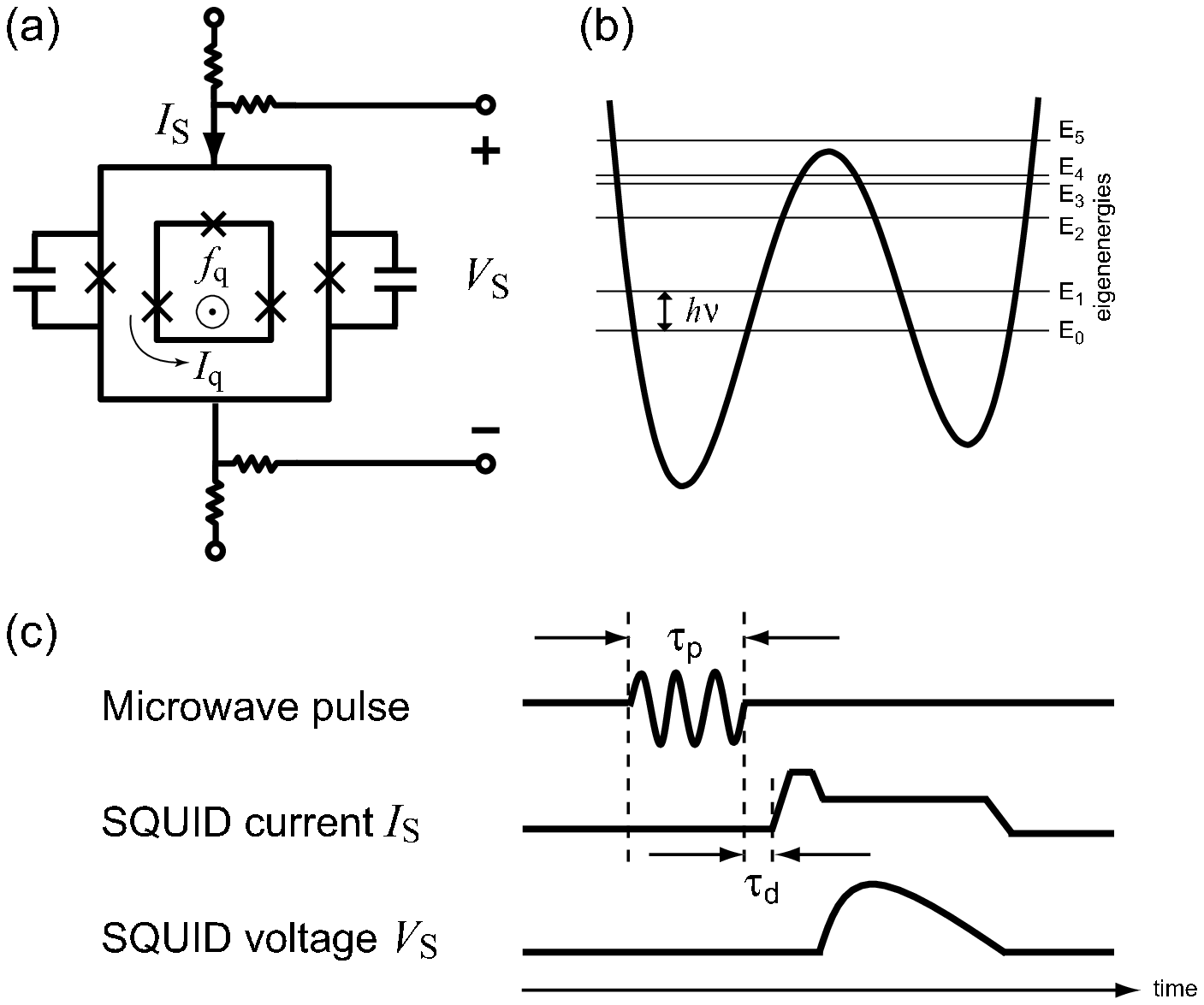}
 \caption{(a) Schematic of the PC qubit surrounded by a readout dc SQUID. ``X'' denotes a
 JTJ. The qubit is biased by a magnetic flux $f_{\text{q}}$, and the circulating current of the qubit is
 $I_{\text{q}}$.
 (b) Simulated double-well potential of the qubit with eigenenergies $E_0 - E_4$ shown. The energy level separation is $h \nu$.
 (c) Time sequence of the microwave pulse, SQUID bias current pulse $I_{\text{S}}$, and SQUID
 voltage response $V_{\text{S}}$ for one measurement trial. The microwave pulse duration is $\tau_{\text{p}}$ and the delay
 time between microwave pulse off and measurement pulse on is $\tau_{\text{d}}$. }
 \label{Fig:fig1}
\end{figure}

\begin{figure}
 \includegraphics{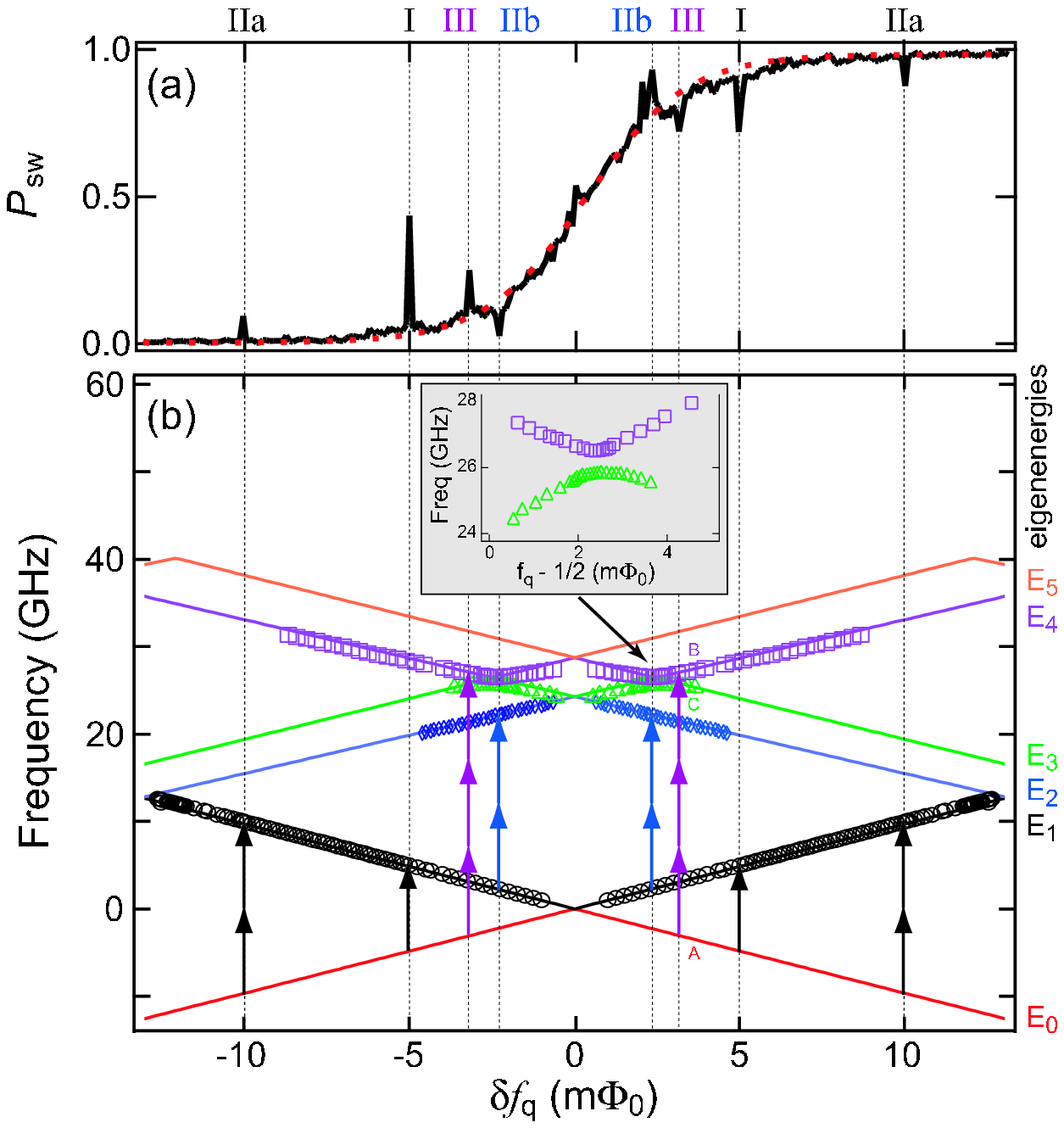}
 \caption{(Color) (a) The
 switching probability $P_{\text{sw}}$ vs. $\delta f_{\text{q}}$ for a
 microwave pulse at frequency $\nu=9.9$~GHz with $\tau_{\text{p}} =
 1$~$\mu$s (solid black line), and thermal distribution in the absence of microwaves (red dashed line).
 Labels I, IIa, IIb, and III indicate single-, two-, and three-photon
 resonances.
 (b) Measured (markers) and fitted (lines) multi-level
 energy structure of the PC qubit. Eigenenergies $E_0 \ldots E_5$ are shown at right.
 A, B, and C indicate levels represented by the three-level model (see text). Inset: zoom-in of the avoided
 crossing between levels $E_3$ and $E_4$.}
  \label{Fig:fig2}
\end{figure}

A PC qubit is a superconducting loop interrupted by three
under-damped Josephson tunnel junctions (JTJs) [see
Fig.~\ref{Fig:fig1}(a)]. Two JTJs are designed to have the same
critical current $I_{\text{c}}$; the third is $ \alpha
I_{\text{c}}$. For $0.5 < \alpha < 1$ and with an externally
applied magnetic flux $f_{\text{q}} \approx \Phi_0/2$ ($\Phi_0
\equiv h/2e$ is the flux quantum), the system is analogous to a
particle in a two-dimensional double-well potential with a
multi-level energy band diagram as simulated (solid lines) in
Fig.~\ref{Fig:fig2}(b). Throughout the paper, the flux will be
parameterized by its detuned valued $\delta f_{\text{q}} \equiv
f_{\text{q}}-\Phi_0/2$. A one-dimensional slice of the double-well
potential biased at a flux detuning $\delta f_{\text{q}} \approx
3$ m$\Phi_0$ is shown in Fig.~\ref{Fig:fig1}(b). Energy levels
with positive and negative slopes correspond to macroscopic
persistent currents $i_{\text{q}}$ of opposing sign, each
associated with one of the potential wells~\cite{Orlando99a}. The
single-well states are coupled via the potential barrier; the
aggregate system has eigenstates with eigenenergies shown in
Figs.~\ref{Fig:fig1}(b) and~\ref{Fig:fig2}(b). Varying the flux
$\delta f_{\text{q}}$ tilts the double-well potential and,
thereby, adjusts its eigenstates and energy-levels. Microwaves
with frequency matching the energy level spacing can generate
transitions between the eigenstates of the undriven
system~\cite{Wal00a,Chiorescu03a,Yu02a}.

The Nb PC qubits were fabricated with a Nb trilayer process at MIT
Lincoln Laboratory~\cite{Berggren99a}. A schematic of the qubit
and readout dc SQUID circuit is shown in Fig.~\ref{Fig:fig1}(a).
The inner loop with three JTJs is the PC qubit with a circulating
current $I_{\text{p}} \approx 0.45$ $\mu$A and $\alpha \approx
0.84$. The critical current density is $J_{\text{c}} \approx 160$
$\text{A}/\text{cm}^2$, $E_{\text{J}} \approx 300$ GHz, and
$E_{\text{C}} \approx 0.65$ GHz. The qubit's loop area is $16
\times 16$ $\mu \text{m}^2$, with a self-inductance of
$L_{\text{q}} \approx 30$ pH. The readout SQUID surrounds the
qubit and consists of two JTJs with equal critical current
$I_{\text{c0}} = 2$ $\mu$A. Both JTJs are shunted with a 1-pF
capacitor to lower the SQUID resonant frequency. The SQUID's loop
area is $20 \times 20$ $\mu \text{m}^2$, with a self-inductance of
$L_{\text{SQ}} \approx 60$ pH. The mutual inductance between the
qubit and dc SQUID is $M \approx 25$ pH.

Measurements were performed in a dilution refrigerator. The device
was magnetically shielded by four cryoperm-10 cylinders and a
superconducting can.  All electrical leads were carefully
attenuated and filtered to minimize noise.
For each measurement cycle, we initialized the system by waiting a
sufficiently long time, typically 5 ms. Then, as illustrated in
Fig.~\ref{Fig:fig1}(c), a microwave pulse with duration
$\tau_{\text{p}}$ is applied to drive quantum-state transitions.
After delaying a time interval $\tau_{\text{d}}$, a readout
current pulse $I_{\text{S}}$ is sent to the SQUID, and the voltage
$V_{\text{S}}$ across the SQUID is monitored with a universal
counter. The $I_{\text{S}}$ pulse consists of a 20 ns short pulse,
with 5 ns rising and falling edges, and an 18 $\mu$s trailing
plateau. The SQUID switches to the finite-voltage state depending
on the qubit state and the current $I_{\text{S}}$.  This procedure
is repeated approximately 1000 times, and the SQUID-switching
counts are used to estimate the switching probability
$P_{\text{sw}}$. In the following experiments, $P_{\text{sw}} = 0$
$(1)$ corresponds to states with positively (negatively) sloped
energy bands.

Fig.~\ref{Fig:fig2}(a) shows an example of how $P_{\text{sw}}$
varied as a function of $\delta f_{\text{q}}$ with microwave
irradiation at $\nu = 9.8$ GHz and of duration $\tau_{\text{p}} =
1$ $\mu$s (solid black line). For reference, the expectation value
of the qubit initial state in the absence of microwaves is fitted
with the thermal distribution function (red dashed
line)~\cite{Orlando99a}. In the presence of microwaves, numerous
resonances (both peaks and dips) are observed. The resonances are
labelled I, IIa, IIb, and III at the top of
Fig.~\ref{Fig:fig2}(a), corresponding to the number of photons
involved in each transition. The flux axes in
Figs.~\ref{Fig:fig2}(a) and~\ref{Fig:fig2}(b) are aligned to
coordinate the resonances in the switching probability trace with
their corresponding position in the energy band diagram. The
experimental data points in Fig.~\ref{Fig:fig2}(b) are indicated
by individual markers, the solid lines are the energy bands
obtained from simulating the qubit using the fabrication
parameters listed above~\cite{Orlando99a}, and color identifies
the first six eigenenergies $E_0 \ldots E_5$. The arrows in
Fig.~\ref{Fig:fig2}(b) are each of height $\nu = 9.9$ GHz; the
number of stacked arrows corresponds to the number of photons
involved in a given transition.

The single-photon resonances, label I in Fig.~\ref{Fig:fig2}(a),
are transitions between the two lowest energy states due to the
applied microwave magnetic field at frequency $(E_1 - E_0)/h =
\nu$. For the same microwave frequency and amplitude, two-photon
resonances, label IIa in Fig.~\ref{Fig:fig2}(a), were also
identified between the two lowest energy states at a flux bias
yielding $(E_1 - E_0)/h = 2 \nu$. In addition, a 3-photon
resonance, label III in Fig.~\ref{Fig:fig2}(a), was identified
between the ground and fourth excited state at a flux bias
yielding $(E_4 - E_0)/h = 3\nu$. Sweeping the microwave frequency
varied the flux-distance between resonance pairs with a slope $|
\Delta \nu / \Delta f_{\text{q}}| \approx 1/\text{n} \text{
GHz}/\text{m}\Phi_0$, allowing us to distinguish
$\text{n}=1\ldots3$ photon transitions and, thereby, reconstruct
the energy band diagram. The reconstructed energy band diagram is
consistent with that simulated using device fabrication parameters
[solid lines, Fig.~\ref{Fig:fig2}(b)], and it provides clear
picture of the multi-level energy-band structure of the qubit.

The additional two-photon resonances, label IIb in
Fig.~\ref{Fig:fig2}(a), are transitions between bands $E_1$ and
$E_2$. A non-zero residual population may exist in $E_1$ due to
finite temperature, and so the IIb-type transitions were generally
observed for flux biases at which $E_1$ had non-zero equilibrium
population [dashed red line in Fig.~\ref{Fig:fig2}(a)]. Since
$E_2$ has the same slope and, thus, circulating current as the
ground state energy $E_0$ at these flux biases, $P_{\text{sw}}$
\textit{decreased (increased)} for $\delta f_{\text{q}} < 0$
$(\delta f_{\text{q}} > 0)$. This is in contrast to transitions I,
IIa, and III, for which $P_{\text{sw}}$ increased (decreased),
because the accessed bands $E_1$ and $E_4$ have opposing slope and
circulating current to $E_0$.

At avoided crossings, the energy bands have zero slope, and the
persistent current vanishes. Thus, at {$\delta f_{\text{q}} = 0$},
the SQUID switching-current readout cannot distinguish the $E_0$
and $E_1$ states. However, we estimated the $E_0$ and $E_1$ tunnel
splitting $\Delta_{12} < 0.1$ GHz by fitting to the eigenenergy
difference $E_1 - E_0 = \sqrt{\Delta_{12}^2 + \varepsilon^2}$,
where $\varepsilon$ is the flux-dependent energy difference
between the uncoupled states~\cite{Wal00a}. The simulated energy
band diagram yielded $\Delta_{12} \sim 10$ MHz. The MHz tunneling
splitting indicates that the coupling between the $E_0$ and $E_1$
states is very weak.

We directly observed an avoided crossing with a tunnel splitting
$\Delta_{34} \approx 500$ MHz between energy levels $E_3$ and
$E_4$ using multi-photon transitions  [inset
Fig.~\ref{Fig:fig2}(b)], indicating a superposition between two
macroscopic quantum states. Although the slopes of $E_3$ and $E_4$
are zero at the avoided crossing, their SQUID readout signals are
distinct from that of the ground state, which has finite slope at
this flux bias. By driving transitions from the ground state, we
could directly measure the $E_3 - E_4$ avoided crossing and its
tunneling splitting $\Delta_{34}$.
\begin{figure}
\includegraphics{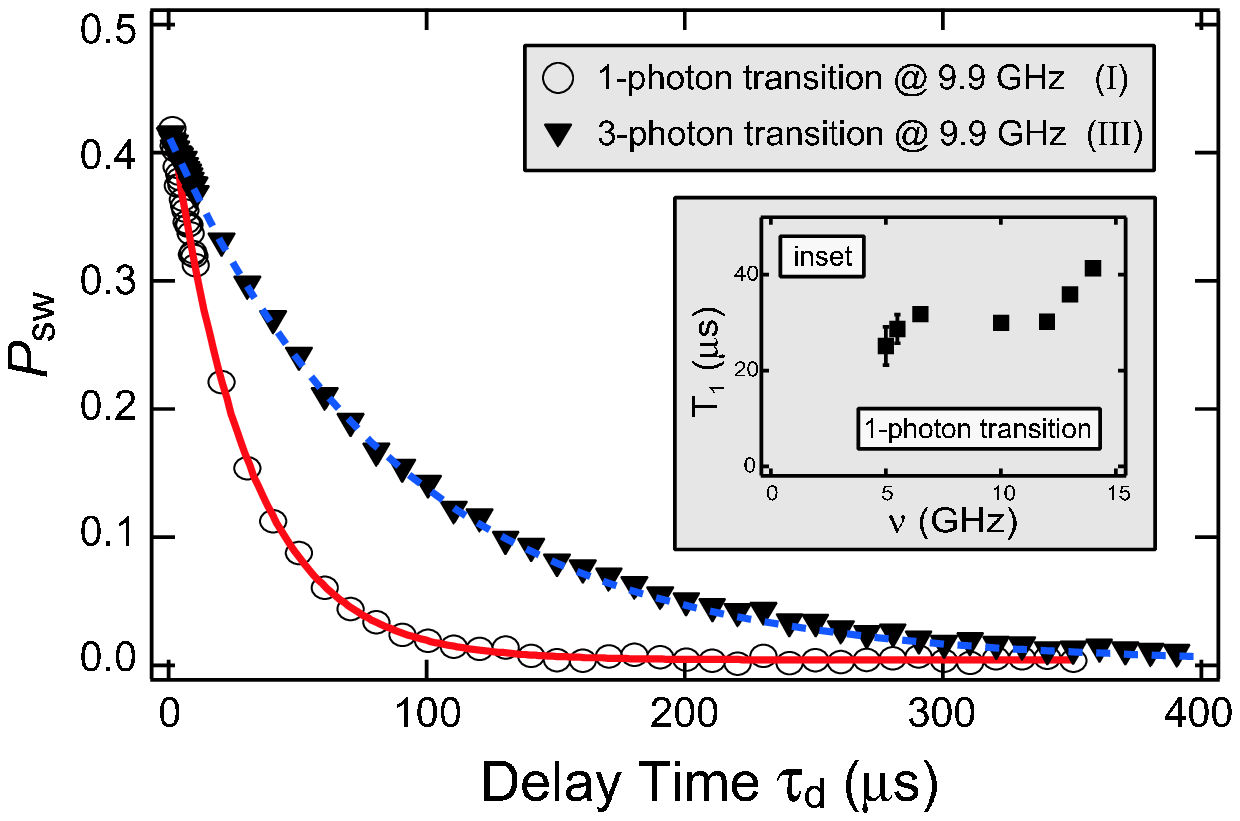}
 \caption{$P_{\text{sw}}$ at $\nu = 9.9$ GHz as a function of
the readout delay time $\tau_{\text{d}}$.  Using a fit exponential
decay, we obtained $T_1 \approx 30 \mu$s for the I transition, and
$T_1 \approx 80$ $\mu$s for the III transition in
Fig.~\ref{Fig:fig2}. Inset: $T_1$ as a function of $\nu$ for the I
transition.}
 \label{Fig:fig3}
\end{figure}

The energy relaxation times $T_1$ between macroscopic quantum
states were measured by driving population to the excited states
($E_1$ and $E_4$ in Fig.~\ref{Fig:fig2}) and monitoring it as a
function of $\tau_{\text{d}}$. Fig.~\ref{Fig:fig3}(a) shows
examples of the population decay; an exponential function fits the
data remarkably well. The obtained relaxation time from $E_1$ is
$T_1 \sim 30$ $\mu$s, and the relaxation time from $E_4$ is even
longer, $T_1 \sim 80$ $\mu$s. Both are consistent with the
independently measured intra-well relaxation time for a similar
qubit design~\cite{Yu04a,Segall03a,Crankshaw03a}. We also
investigated $T_1$ as a function of the energy difference $h \nu$
(inset Fig.~\ref{Fig:fig3}). The experimental $T_1$ tends to
increase with increasing $\nu$, qualitatively agreeing with the
spin-boson bath model~\cite{Leggett87a,Makhlin01a}. However, the
deviation between the data and theory suggests that we may have
other sources and structure to the dissipation. Nevertheless, we
can potentially benefit from this long $T_1$ in future
experiments, such as the three-level flux qubit~\cite{Zhou02a} and
superconductive electromagnetic induced
transparency~\cite{Murali04a}.
\begin{figure}
\includegraphics{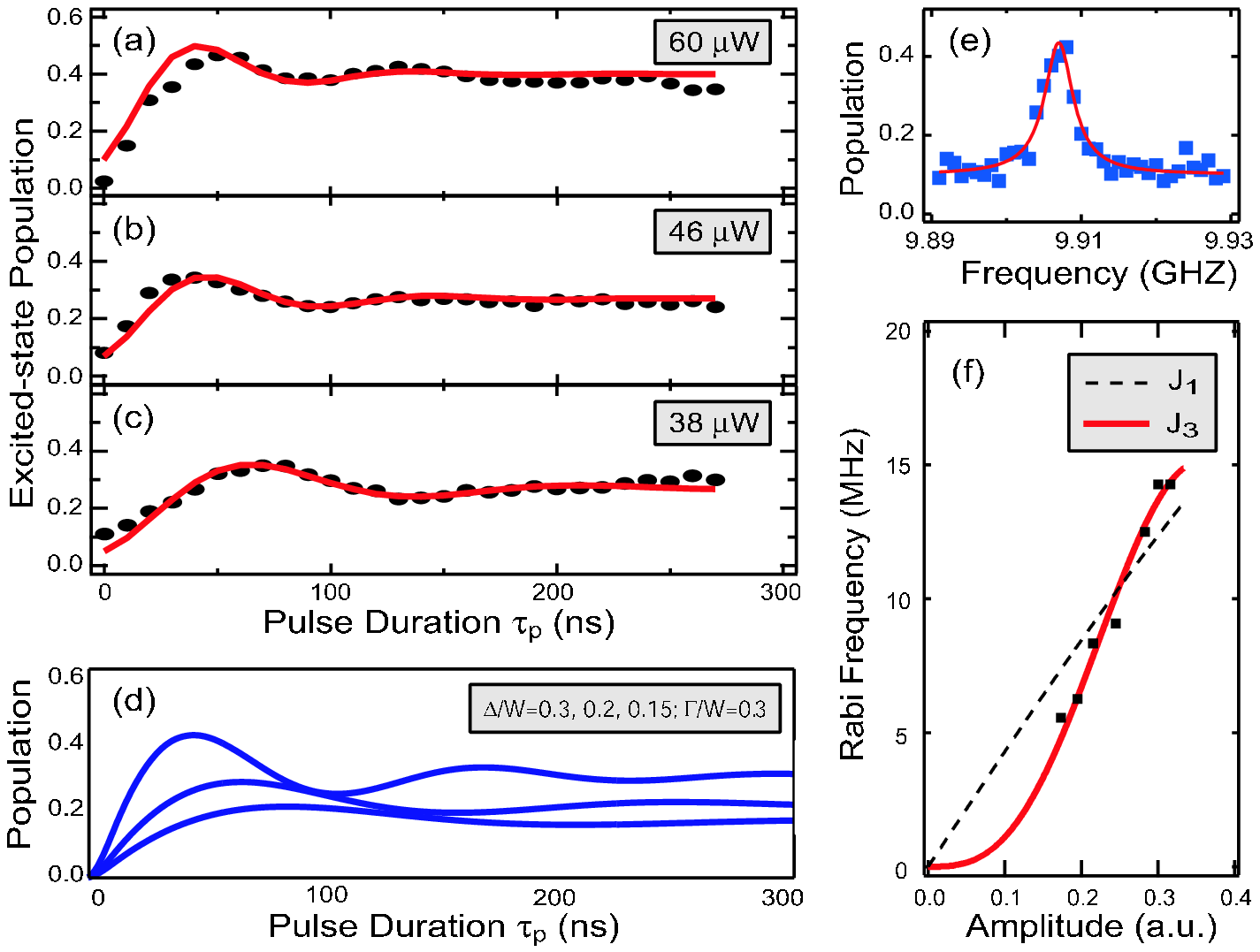}
 \caption{(a)-(c) The amplitude of the 3-photon resonant peaks
(III in Fig.~\ref{Fig:fig2}) vs. the microwave pulse duration
$\tau_{\text{p}}$ for $\nu= 9.90779$ GHz at different powers. The
population oscillations with different frequencies are clearly
observed. (d) Coherent oscillations simulated using a three-level
model. (e) Lorentzian-fitted (red line) resonance peak data (blue
markers) with FWHM $\sim 5$ MHz. The corresponding dephasing time
is $\sim 100$ ns. (f) The Rabi frequency as a function of the
microwave amplitude. The solid (dashed) line is a best fit to a
third-order (first-order) Bessel function of the first kind.}
 \label{Fig:fig4}
\end{figure}

Although the coupling between the ground and first excited states
proved too weak to see definitive Rabi oscillations, we were able
to observe oscillations in $P_{\text{sw}}$ for transitions between
the ground and fourth excited states, since the presence of the
avoided crossing (Fig.~\ref{Fig:fig2} inset) gives a larger
overlap between these states. We biased $\delta f_{\text{q}}$ at
the 3-photon resonant peak (III in Fig.~\ref{Fig:fig2}) and
measured the population in the excited state $P_{\text{sw}}$ as a
function of microwave pulse duration time $\tau_{\text{p}}$, shown
in Fig.~\ref{Fig:fig4}(a)-(c). The population $P_{\text{sw}}$
oscillates as a function of $\tau_{\text{p}}$, and the oscillation
frequency changes with the microwave amplitude. The Rabi decay
time is approximately 50 ns, although the oscillation visibilities
are only approximately $20 \%$. In Fig. 4(f), the oscillation
frequencies were plotted as a function of the microwave amplitude.
Coherent oscillations were also observed at several frequencies in
the range $9 < \nu < 10$ GHz near the avoided crossing (not
shown).

The oscillations in Fig.~\ref{Fig:fig4} arise from multi-photon
transitions in a multi-level system. We modelled the multi-level,
multi-photon transition using a three-level system in which
coherent oscillations are driven between levels $A$ and $B$, and
level $B$ is taken to be strongly coupled to a third level $C$
[see Fig.~\ref{Fig:fig2}(b)]. Fig.~\ref{Fig:fig4}(d) shows
coherent oscillations simulated using this model, assuming
coupling strengths and decoherence parameters consistent with our
qubit. Details of this model will be presented elsewhere, and we
only state the qualitative results here: driving levels $A$ and
$B$ modifies the coupling between $B$ and $C$, and this
modification is dependent upon the driving amplitude. The presence
of the third level $C$ results in an amplitude-dependent
visibility that vanishes for small driving amplitudes, while, for
larger amplitudes, the power-dependent coupling to the third level
$C$ effectively detunes the $A$ - $B$ transition. The net result
is a limited window of driving amplitudes over which multi-photon
transitions can be clearly observed in a multi-level system.

In the limit that the coupling to $C$ is eliminated, the above
model reduces to the n-photon, two-level case, in which the Rabi
frequency will vary as $J_{\text{n}}(x)$, where $J_{\text{n}}$ is
the n-th order Bessel function of the first kind and
$x=eV_{\text{ac}}/h \nu$ is proportional to the microwave driving
amplitude $V_{\text{ac}}$~\cite{Nakamura}. The experimental
amplitude dependence of the Rabi frequency in
Fig.~\ref{Fig:fig4}(f) is reasonably fit by a third-order Bessel
function over the limited range of amplitudes, with the reduced
visibilities observed experimentally in Fig.~\ref{Fig:fig4}(a)-(c)
and simulated in Fig.~\ref{Fig:fig4}(d). Despite the
fringe-contrast reduction due to the multiple levels, the $\sim
50$ ns decoherence time of our Nb PC qubit is comparable to those
of the qubits fabricated with other superconducting
materials~\cite{Vion02a,Yu02a,Martinis02a,Chiorescu03a}.

In summary, we demonstrated multi-level, multi-photon coherent
dynamics in a niobium PC qubit. From the single- and multi-photon
resonances, we extracted a multi-level energy band diagram. We
directly measured an avoided crossing, and we demonstrated
three-photon coherent oscillations between the ground and fourth
excited state in a multi-level environment.

We gratefully acknowledge E. Macedo, T. Weir, R. Slattery, G.
Fitch, and D. Landers for technical assistance with the device
fabrication, packaging, and testing at MIT Lincoln Laboratory. We
also thank D. Nakada, J Habif, and D. Berns for technical
assistance at the MIT campus, and D. Cory and S. Lloyd for helpful
discussions. This work was supported by AFOSR (F49620-01-1-0457)
under the DURINT program. The work at Lincoln Laboratory was
sponsored by the US DoD under Air Force Contract No.
F19628-00-C-0002.

\bibliographystyle{Will}

\begin{thebibliography}{10}
\bibitem{macro_meso_note}
We note an additional distinction between superconducting flux and
charge qubits made by some authors, e.g.,
Leggett~\cite{Leggett87a} and Averin~\cite{Averin99a}, with regard
to the number of microscopic degrees of freedom involved in the
coherent evolution. In flux qubits, this number is macroscopically
large, warranting the name ``macroscopic coherence.'' In charge
qubits, in contrast, most of the charge degrees of freedom, except
a few, do not participate in the coherent evolution, and thus the
aforementioned authors use the ``mesoscopic'' rather than the
``macroscopic'' attribution.
\end{thebibliography}

\end{document}